\documentclass[preprint,authoryear,12pt]{elsarticle}

\usepackage{graphicx}
\usepackage{amssymb}

\usepackage{subcaption}
\usepackage{amsmath}
\usepackage{url}
\usepackage{color}

\journal{Advances in Space Research}

\begin{document}

%%%%%%%%%%%%%%%%%%%%%%%%%%%%%%%%%%%%%%%%%%%%%%%%%%%%%%%%%%%%%%%%%%%%%%%%%%%%%
%% Frontmatter
\begin{frontmatter}


% Use the tnoteref command within \title and fnref within \author or \address for footnotes;
% use the corref command within \author for corresponding author footnotes;
% \author{Name\corref{cor1}\fnref{label2}}
% \address{Address\fnref{label3}}

\title{The gravitational redshift monitored with RadioAstron from near Earth up to 350,000 km}

\author[add1]{N. V. Nunes\corref{cor}}
\ead{nvnunes@yorku.ca}
\cortext[cor]{Corresponding author}
\author[add1]{N. Bartel}
\author[add1,add2]{M. F. Bietenholz}
\address[add1]{York University, 4700 Keele St, Toronto, ON M3J 1P3, Canada}
\address[add2]{South African Radio Astronomy Observatory, P.O. Box 443, Krugersdorp 1740, South Africa}

\author[add3]{M. V. Zakhvatkin}
\address[add3]{Keldysh Institute for Applied Mathematics, Russian Academy of Sciences, Miusskaya sq. 4, 125047 Moscow, Russia}

\author[add4,add5,add6]{D. A. Litvinov}
\author[add4]{V. N. Rudenko}
\address[add4]{Sternberg Astronomical Institute, Lomonosov Moscow State University, Universitetsky pr. 13, 119234 Moscow, Russia}
\address[add5]{Astro Space Center, Lebedev Physical Institute, Profsoyuznaya 84/32, 117997 Moscow, Russia}
\address[add6]{Bauman Moscow State Technical University, 2-ya Baumanskaya 5, 105005 Moscow, Russia}

\author[add7,add8]{L. I. Gurvits}
\author[add8]{G. Granato}
\author[add8]{D. Dirkx}
\address[add7]{Joint Institute for VLBI ERIC, Oude Hoogeveensedijk 4, 7991 PD Dwingeloo, Netherlands}
\address[add8]{Delft University of Technology, Mekelweg 5, 2628 CD Delft, Netherlands}

\begin{abstract}
We report on our efforts to test the Einstein Equivalence Principle by measuring the gravitational redshift with the VLBI spacecraft RadioAstron, in an eccentric orbit around Earth with geocentric distances as small as $\sim$ 7,000 km and up to 350,000 km. The spacecraft and its ground stations are each equipped with stable hydrogen maser frequency standards, and measurements of the redshifted downlink carrier frequencies were obtained at both 8.4 and 15 GHz between 2012 and 2017. Over the course of the $\sim$ 9 d orbit, the gravitational redshift between the spacecraft and the ground stations varies between $6.8 \times 10^{-10}$ and $0.6 \times 10^{-10}$. Since the clock offset between the masers is difficult to estimate independently of the gravitational redshift, only the variation of the gravitational redshift is considered for this analysis. We obtain a preliminary estimate of the fractional deviation of the gravitational redshift from prediction of $\epsilon = -0.016 \pm 0.003_{\rm stat} \pm 0.030_{\rm syst}$ with the systematic uncertainty likely being dominated by unmodelled effects including the error in accounting for the non-relativistic Doppler shift. This result is consistent with zero within the uncertainties. For the first time, the gravitational redshift has been probed over such large distances in the vicinity of Earth. About three orders of magnitude more accurate estimates may be possible with RadioAstron using existing data from dedicated interleaved observations combining uplink and downlink modes of operation.
\end{abstract}

\begin{keyword}
%first keyword \sep second keyword \sep more keywords
Test of general relativity; Einstein Equivalence Principle; RadioAstron; space VLBI
\end{keyword}

\end{frontmatter}

\parindent=0.5 cm

%%%%%%%%%%%%%%%%%%%%%%%%%%%%%%%%%%%%%%%%%%%%%%%%%%%%%%%%%%%%%%%%%%%%%%%%%%%%%
%% Main text

\section{Introduction}

The incompatibility of general relativity and quantum theory is a fundamental problem in our understanding of the physical world. The Einstein Equivalence Principle (EEP) is a cornerstone of general relativity, and it leads to there being a gravitational redshift \citep{Will93}. Specifically, local position invariance requires that time flows slower for an observer close to a massive body than for one farther away. Similarly, an electromagnetic wave is seen to be redshifted when its source is located close to a massive body in comparison to when the source is located far away from it. Accurate measurements of the gravitational redshift are thus a way of verifying EEP and, in turn, general relativity. Comparisons with predictions are therefore of prime importance.  

A milestone in measuring the gravitational redshift was reached with the Gravity Probe A (GP-A) mission in 1976. A Scout rocket lifted a spacecraft with a hydrogen maser frequency standard on board to an altitude of 10,000 km. The frequency from the onboard maser was compared with that from a maser on the ground. The resulting gravitational redshift measurement was consistent with prediction and had a fractional uncertainty of $1.4 \times 10^{-4}$ \citep{VesLev79,Ves89}. Recently, the two teams working on the GREAT project, using data from the Galileo 5 and 6 satellites erroneously launched into slightly elliptical orbits with an eccentricity of 0.166 and semi-major axis of $\sim$ 27,980 km, have published results that reduce this uncertainty by more than a factor of 4 \citep{Herr18} and 5.6 \citep{Del18} respectively. In the future, the ACES experiment, on board the International Space Station in a nearly circular orbit at an altitude of $\sim$ 400 km, is expected to reduce the uncertainty down to the $10^{-6}$ level \citep{Mey18}.

We conducted a similar experiment to GP-A in conjunction with the Russian-led international space Very Long Baseline Interferometry (space VLBI) mission RadioAstron \citep{Kar13}. The RadioAstron spacecraft was launched on 2011 July 18 into a highly elliptical orbit around Earth with a perigee as small as $\sim$ 7,000 km and an apogee as large as $\sim$ 350,000 km from the geocenter.  Since early on, the spacecraft has been used for astrophysical space VLBI observations of compact radio sources. The spacecraft has a hydrogen maser frequency standard (H-maser) on board that operated until mid-2017. Ground testing of this H-maser showed an Allan deviation of $2 \times 10^{-15}$ for an averaging time of 1 hour. Prior to the failure of the onboard H-maser, the downlink carrier frequencies of 8.4 and 15 GHz, locked to the frequency of the maser, were recorded at ground stations, also equipped with H-masers, during science observations. The ground stations for RadioAstron are the 26-m Pushchino tracking station in Russia (PU) and, from 2013 onwards, the 43-m Green Bank tracking station at the West Virginia site of the National Radio Astronomy Observatory in the USA (GB). We refer to this space-to-ground mode of operation as the 1-way mode. During the $\sim$ 9 d elliptical orbits, the spacecraft travels through the gravitational potential of Earth, which according to the EEP, should cause a varying gravitational redshift on the downlink signals with a relative frequency shift predicted to vary between $6.8 \times 10^{-10}$ and $\sim 0.6 \times 10^{-10}$. In order to measure the gravitational redshift, various effects need to be subtracted from the observed frequencies, the largest being the non-relativistic Doppler shift. This must be computed using the orbit of the spacecraft. 

\begin{figure}[!t]
\begin{center}
\begin{subfigure}{0.5\textwidth}
	\centering
	\includegraphics*[width=5cm]{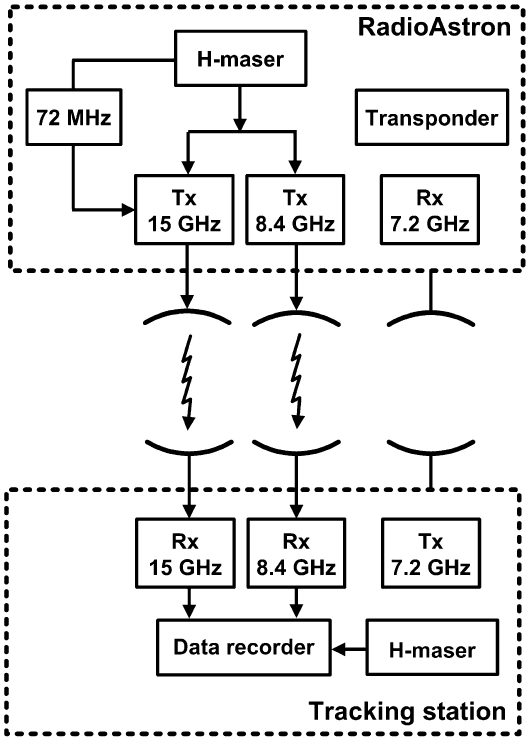}
	\caption{}
\end{subfigure}%
\begin{subfigure}{0.5\textwidth}
	\centering
	\includegraphics*[width=5cm]{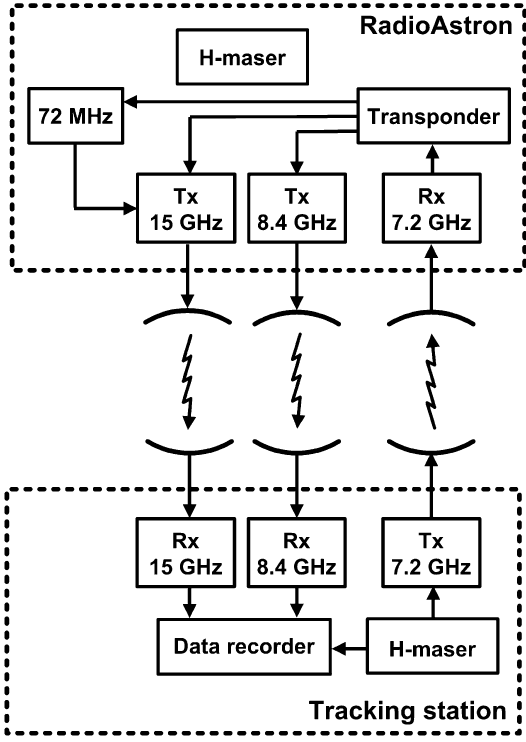}
	\caption{}
\end{subfigure}%
\end{center}
\caption{Block diagram of RadioAstron's modes of operation \citep{Lit18}. In the 1-way mode (a), the downlink carrier frequencies from the spacecraft are locked to the onboard H-maser. Measuring these frequencies at the station, also equipped with an H-maser, allows the gravitational redshift to be monitored. In the 2-way mode (b), the onboard H-maser is not used. A combination of these modes is used to mostly cancel the non-relativistic Doppler shift and the effect of the troposphere.}
\label{figure1}
\end{figure}

In addition to the 1-way mode, the spacecraft  also operates in a 2-way mode where the downlink signals are phase coherently locked to the uplink carrier frequency from the ground station. Figure \ref{figure1} gives a block diagram for the two modes. In the 2-way mode, the Doppler shift is doubled while the gravitational redshift is cancelled. A combination of both the 1-way and 2-way modes that cancels the non-relativistic Doppler shift but retains the gravitational redshift allows the latter to be measured with much higher accuracy. This combination is similar to the GP-A mode of operation. The 1-way mode and the combination of the 1-way and 2-way modes each have advantages and disadvantages. The 1-way mode is compatible with normal science observations and thus allowed almost continuous daily monitoring of the predicted varying gravitational redshift for $\sim$ 5 yr, albeit with relatively low accuracy. Switching between 1-way and 2-way modes, however, is not compatible with normal science observations and requires dedicated interleaved observations and so the varying gravitational redshift was monitored for relatively short periods, but with expected higher accuracy. Here we report on our analysis of 1-way mode measurements recorded at the ground stations and give preliminary results. 

\section{Modelling the 1-way data}
The fractional frequency shift, $\Delta f_{\rm obs} / f$, measured at a station is given by the equation:

\begin{equation}
\begin{split}
\frac{\Delta f_{\rm obs}}{f} =-\frac{\dot{D}}{c} - \frac{{v_s}^2 - {v_e}^2}{2c^2}+ \frac{(\mathbf {v_s} \cdot \mathbf n)^2 -(\mathbf{ v_e} \cdot \mathbf n) \cdot (\mathbf {v_s} \cdot \mathbf n)}{c^2} \\
+\frac{\Delta f_{\rm grav}}{f} +\frac{\Delta f_{\rm clock}}{f} +\frac{\Delta f_{\rm trop}}{f} + \frac{\Delta f_{\rm fine}}{f}+O(\frac{v^3}{c^3})
\end{split}
\label{eq:1}
\end{equation}

\noindent as described in \citet{Bir14}, see also \citet{Saz10}. The first term is the non-relativistic Doppler shift, of order $10^{-5}$, where $\dot{D}$ is the range rate or radial velocity between the spacecraft and ground station and $c$ is the speed of light. The second and third terms are the relativistic contributions to the Doppler shift of order $10^{-10}$, where $\mathbf{v_e}$ and $\mathbf{v_s}$, respectively, are the velocity vectors of the station and spacecraft in the J2000 reference frame and $\mathbf n$ is the unit vector in the direction opposite to that of signal propagation. The fourth term is the observed gravitational redshift between the ground station and spacecraft and is of order $10^{-10}$. The fifth term, $\Delta f_{\rm clock} / f$, of order $10^{-11}$, is due to the clock offset between the H-masers which drift relative to each other over time, and includes both a constant clock offset, $y_0$, and to first order a linear clock drift, $y_1$, on the order of $10^{-14}$/d with $\Delta f_{\rm clock} / f = y_0+y_1t$. The sixth term, $\Delta f_{\rm trop} / f$, of order up to $10^{-11}$, is due to the troposphere. Effects of order $10^{-13}$ and smaller arising from the ionosphere, antenna phase center motion effects and the gravitational effects of the Moon and Sun are grouped together in $\Delta f_{\rm fine} / f$. Finally, $O(v^3 / c^3)$, of order $10^{-15}$, consists of third order or higher terms that need not be considered given the accuracy achievable with our analysis.

\begin{figure}[!t]
\begin{center}
\includegraphics*[width=10cm]{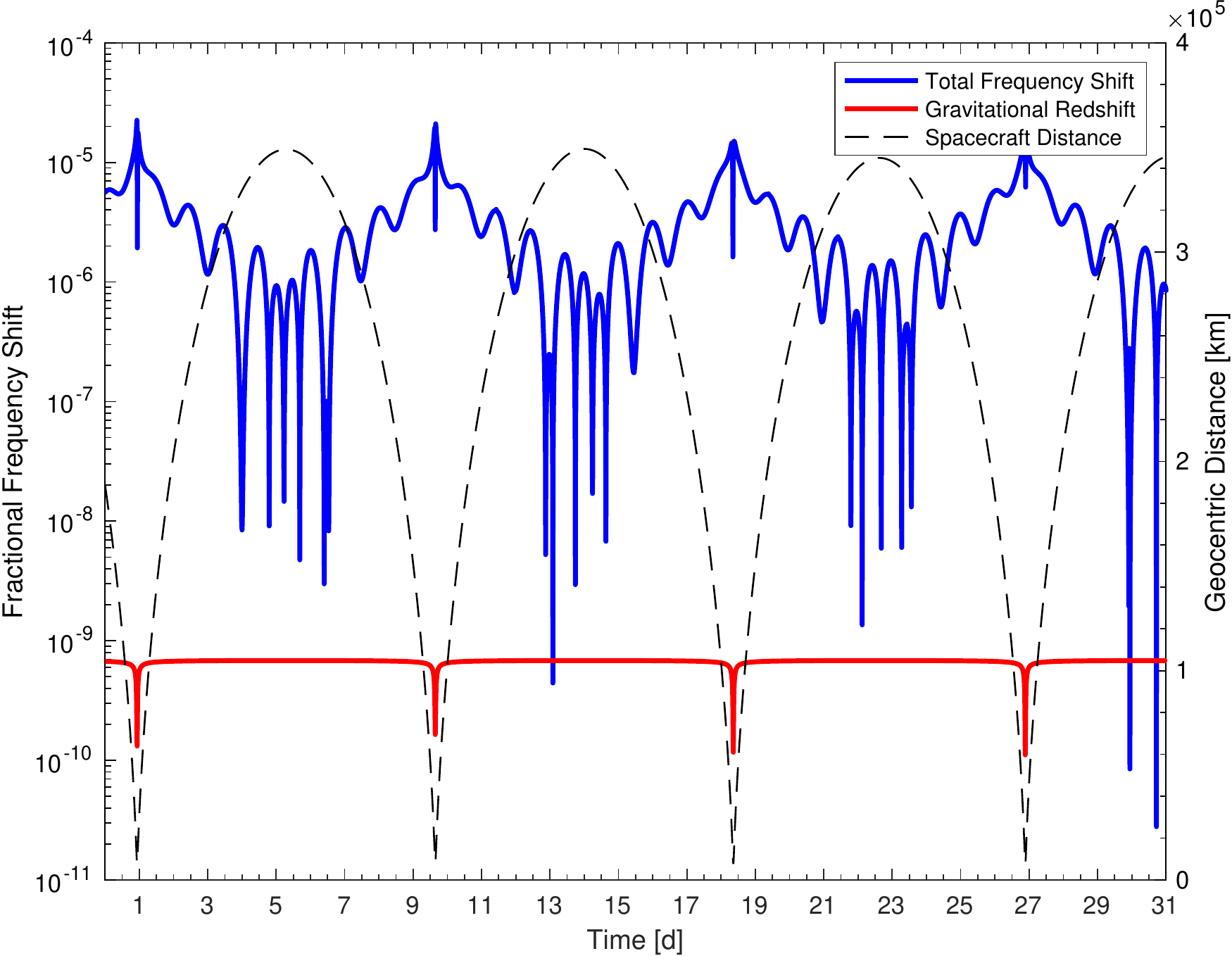}
\end{center}
\caption{A plot of the predicted total fractional frequency shift between the spacecraft and the GB ground station for a number of $\sim$ 9 d orbits, dominated by the non-relativistic Doppler shift, $-\dot{D}/c$, along with the much smaller gravitational redshift, $\Delta f_{\rm grav} / f$, with absolute values plotted against the logarithmic scale on the left side. In addition, the geocentric distance of the spacecraft, $|\mathbf {r_s}|$, is plotted with the scale on the right side. Time is given in days since 2014 January 1.}
\label{figure2}
\end{figure}

In Figure \ref{figure2}, we plot the total frequency shift and predicted gravitational redshift for the month of January in 2014 together with the geocentric distance, $|\mathbf{r_s}|$, of the spacecraft. The non-relativistic Doppler shift, which dominates the total frequency shift, varies due to the motion of the spacecraft and Earth's rotation. The largest fractional frequency shifts are more than $10^{-5}$. The gravitational redshift is about $10^4$ times smaller. Due to the large eccentricity of the orbit, the spacecraft spends much more time near apogee, where the gravitational redshift is near its maximum value of $\sim 6.8 \times 10^{-10}$. Near perigee, the gravitational redshift changes rapidly, reaching a minimum that depends on the orbit, which is constantly evolving, but can be as small as $\sim 0.6 \times 10^{-10}$.

\section{Feasibility of determining the violation parameter $\epsilon$}

A violation of the predicted gravitational redshift can be parametrized by introducing a violation parameter, $\epsilon$, defined as:

\begin{equation}
{%
z_{\rm obs}= (1+\epsilon)z
}
\label{eq:2}
\end{equation}

%\noindent where $z \equiv \frac{GM_{\bigoplus}}{c^{2}}(\frac{1}{r_e}-\frac{1}{r_s})$, $G$ being the gravitational constant, $M_{\bigoplus}$ the mass of Earth, $r_e$ and $r_s$ the respective distances of the station and spacecraft from Earth's center of mass in the J2000 reference frame. 

\noindent  where $z \equiv \Delta U / c^{2}$ with $\Delta U$ being the difference in Earth's gravitational potential between the ground station and the spacecraft. The parameter $\epsilon$ is what we seek to measure with $\epsilon=0$ corresponding to EEP being valid. To look for a possible violation using measured fractional frequency shifts, the state parameters for the spacecraft must be known with sufficient accuracy to isolate the gravitational redshift from other effects, particularly the much larger non-relativistic Doppler effect. Orbit determination for the RadioAstron mission is done using a variety of sources of data, among them the same frequency observations made at the PU and GB ground stations used in this analysis \citep{Zak13, Zak18}. The position and velocity of the spacecraft are determined using a least-squares minimization in which no violation is assumed. We conducted a covariance analysis, done independently by two groups within our collaboration, to investigate whether the determination of $\epsilon$ would be biased by this assumption. No significant correlation between $\epsilon$ and the state parameters was found when they are estimated concurrently. However, as was expected, a significant correlation was found between $\epsilon$ and the constant clock offset, $y_0$. Since the constant clock offset is difficult to measure independently of the gravitational redshift, it is necessary to modify the approach to determining $\epsilon$ by instead looking at the variation of the gravitational redshift over the course of an orbit.

We define a new quantity, the observed biased gravitational redshift as: 
\begin{equation}
\begin{aligned}
z_{\rm obs}^{\rm b} & =\frac{\Delta f_{\rm obs}}{f} - \frac{\Delta f_{\rm doppler}}{f} -\frac{\Delta f_{\rm trop}}{f} \\
& = \frac{\Delta f_{\rm grav}}{f}+\frac{\Delta f_{\rm clock}}{f}
\end{aligned}
\label{eq:3}
\end{equation}

\noindent obtained by subtracting the modelled Doppler effects (first three terms in Equation \ref{eq:1}) and the modelled tropospheric effects from $\Delta f_{\rm obs} / f$, leaving the gravitational redshift biased by the clock offset. The tropospheric effect is computed using a tropospheric delay model based on seasonal averages \citep{Col99}. Contributions from effects on the order of $10^{-13}$ or smaller including the ionosphere and the tidal effects of the Moon ($\sim 10^{-13}$) and Sun ($\sim 10^{-14})$ are too small compared to the accuracy achievable with our present analysis and are ignored. The observed biased gravitational redshift can thus be parametrized as:

\begin{equation}
{%
z_{\rm obs}^{\rm b} = (1+\epsilon)z + y_0+ y_1t
}
\label{eq:4}
\end{equation}

\noindent By taking the difference between pairs of observations, the constant in the clock offset is removed:

\begin{equation}
{%
\Delta z_{\rm obs}^{\rm b} = (1+\epsilon) \Delta z + y_1 \Delta t
}
\label{eq:5}
\end{equation}

\noindent However, the expected uncertainty in the estimate of $\epsilon$ determined in this way increases by more than an order of magnitude as the variation of the gravitational redshift, $\Delta z$, is at least an order of magnitude smaller, on average, than the value of the gravitational redshift, $z$. 

\begin{figure}[!t]
\begin{center}
\begin{subfigure}{0.75\textwidth}
\vspace{-7pt}
\includegraphics*[width=10cm]{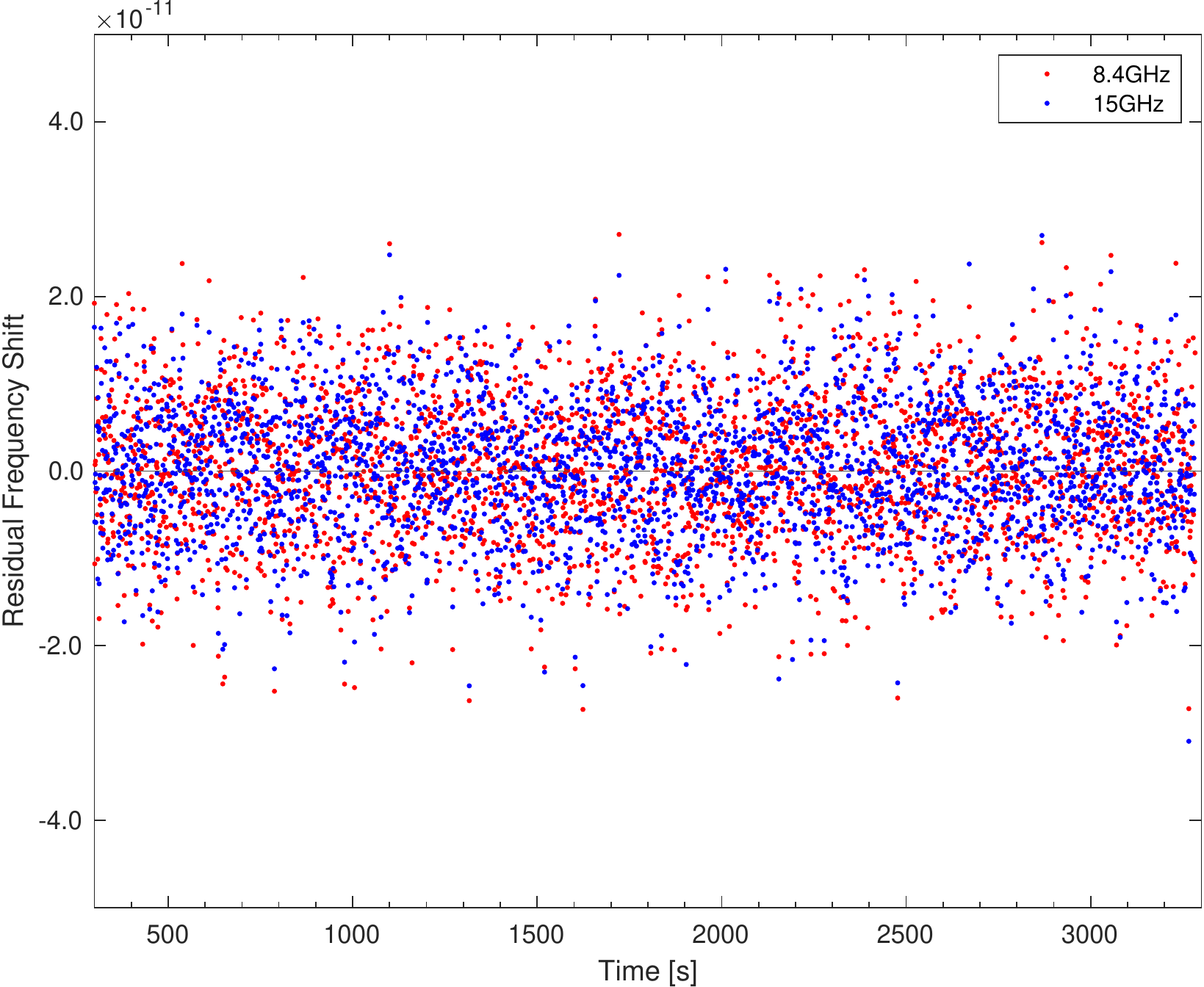}
\end{subfigure}%
\begin{subfigure}{0.25\textwidth}
\includegraphics*[width=7.9cm,angle=-90]{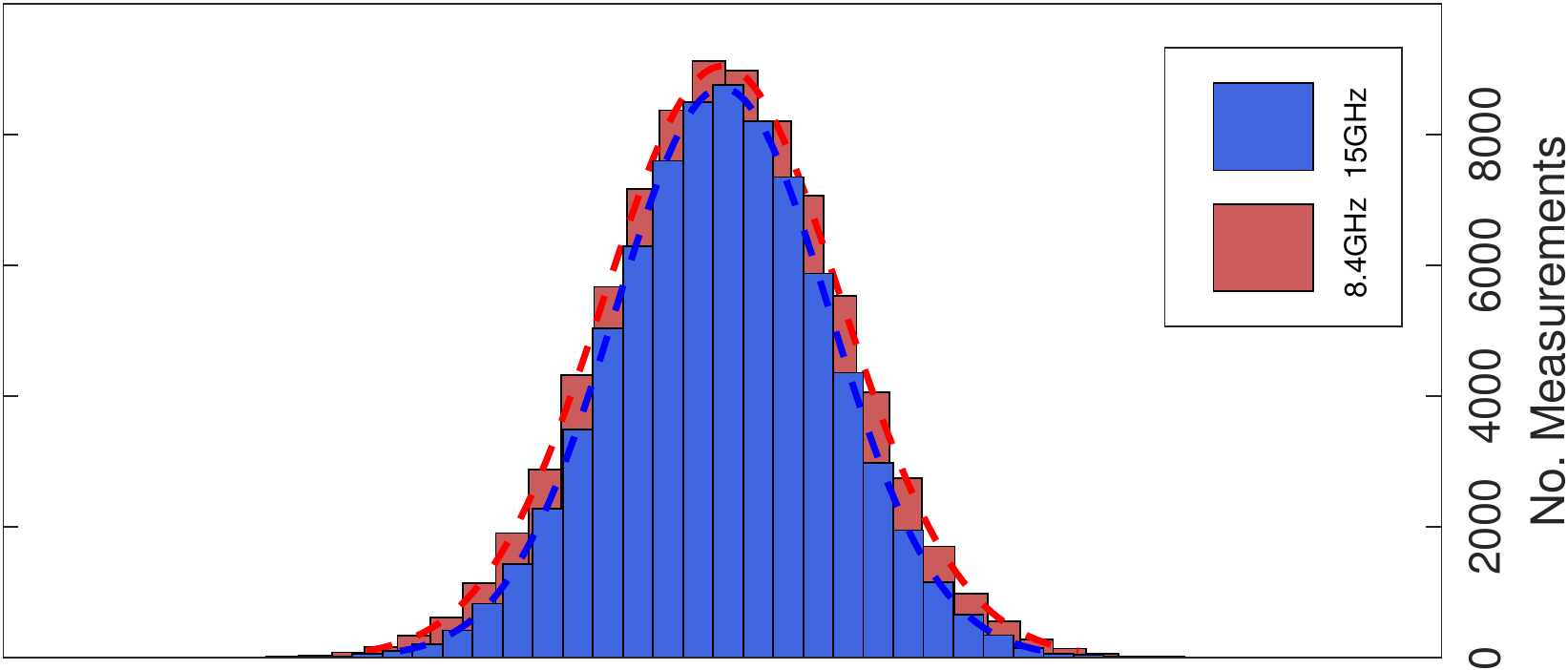}
\end{subfigure}
\end{center}
\caption{A sample of the $\sim$ 100,000 residuals from fitting the frequency measurements at 8.4 and 15 GHz with a polynomial for a single session of $\sim$ 1 h at Pushchino. The right panel shows the distributions of all the residuals at each downlink carrier frequency with almost complete overlap and with the dashed lines being Gaussian fits. } 
\label{figure3}
\end{figure}

\section{Observations and raw data}

About 3,900 sessions of downlink carrier frequency measurements were performed between 2012 and 2017, about 2,700 at PU and the remaining 1,200 at GB.  Each session is about 1 h long and most have data for both the downlink carrier frequencies of 8.4 and 15 GHz. During each session, frequencies were recorded every 40 ms, resulting in over 100,000 frequency measurements at each of the two downlink carrier frequencies. A total of 850 million frequency measurements were obtained. At 40 ms intervals, frequency measurements are affected by Gaussian noise with a fractional frequency shift on the order of $10^{-11}$ and must be smoothed in time. The frequency measurements from each session are fit with a polynomial that is used to arrive at a single frequency measurement at the mid-point of the session. Figure \ref{figure3} displays the residuals from such a fit for one session.  Sessions after June 2017 are excluded as at that time the onboard H-maser began to show signs of failure prior to running out of hydrogen later in the summer. GB sessions in 2013 when the station was temporarily on a Rubidium standard are also excluded. In addition, $\sim$ 12\% of the remaining sessions where residuals aren't sufficiently Gaussian distributed are also excluded. These additional excluded sessions are spread somewhat evenly over the full 5 years for both stations.

\section{Data analysis}
\begin{figure}[!p]
\begin{center}
\begin{subfigure}{0.5\textwidth}
	\includegraphics*[width=6.5cm]{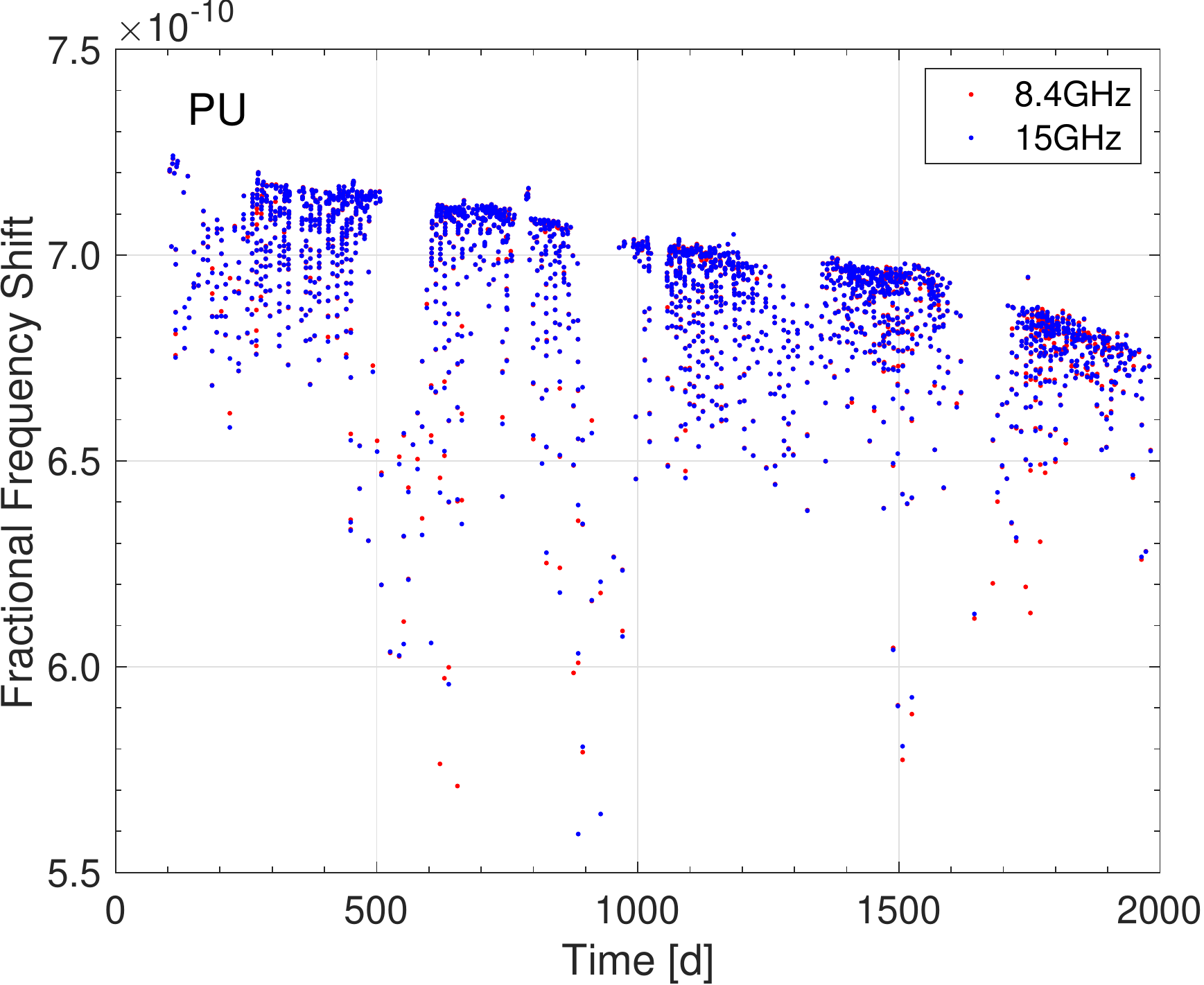}
	\caption{}
\end{subfigure}%
\begin{subfigure}{0.5\textwidth}
	\includegraphics*[width=6.5cm]{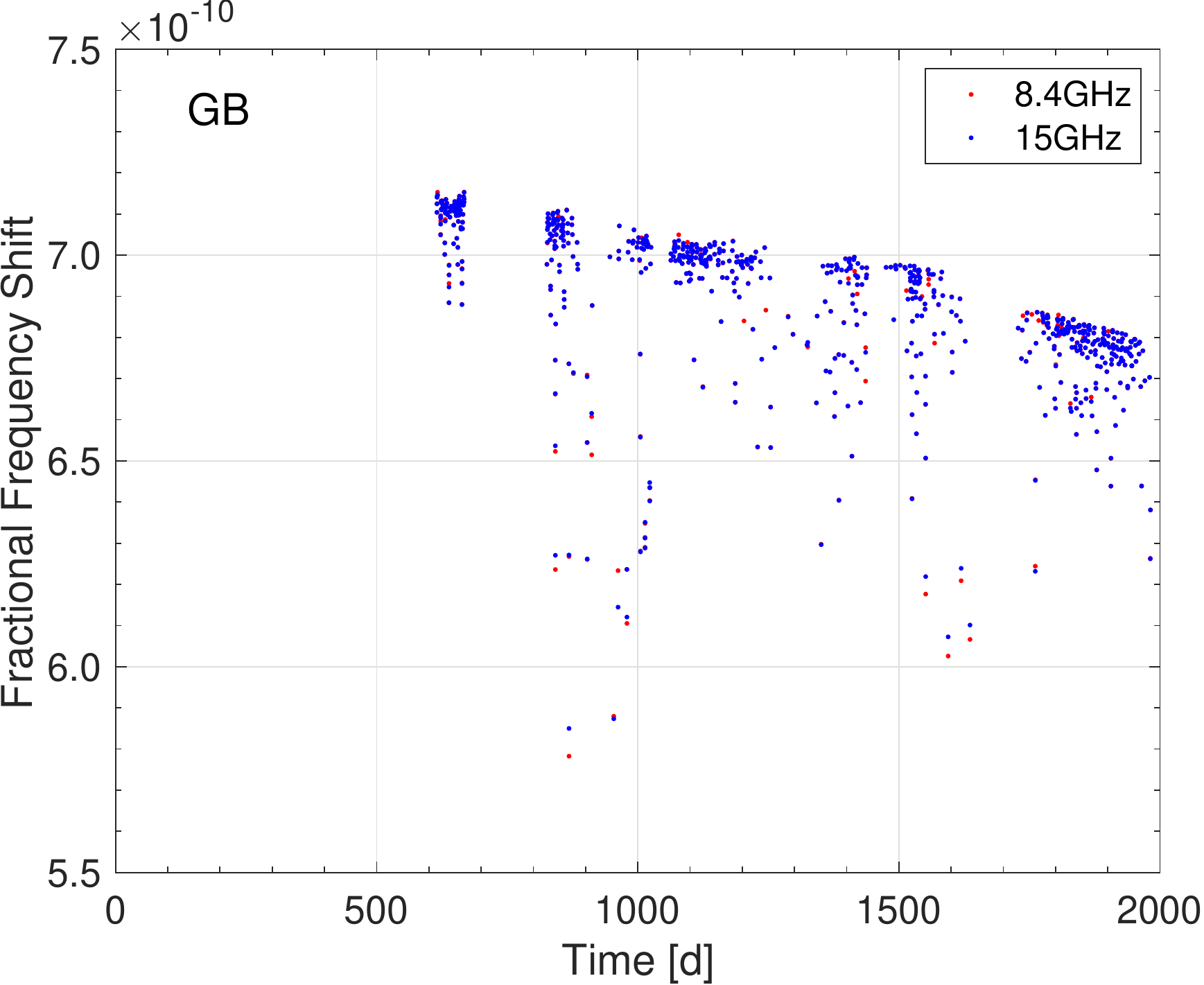}
	\caption{}
\end{subfigure}
\begin{subfigure}{0.5\textwidth}
	\includegraphics*[width=6.5cm]{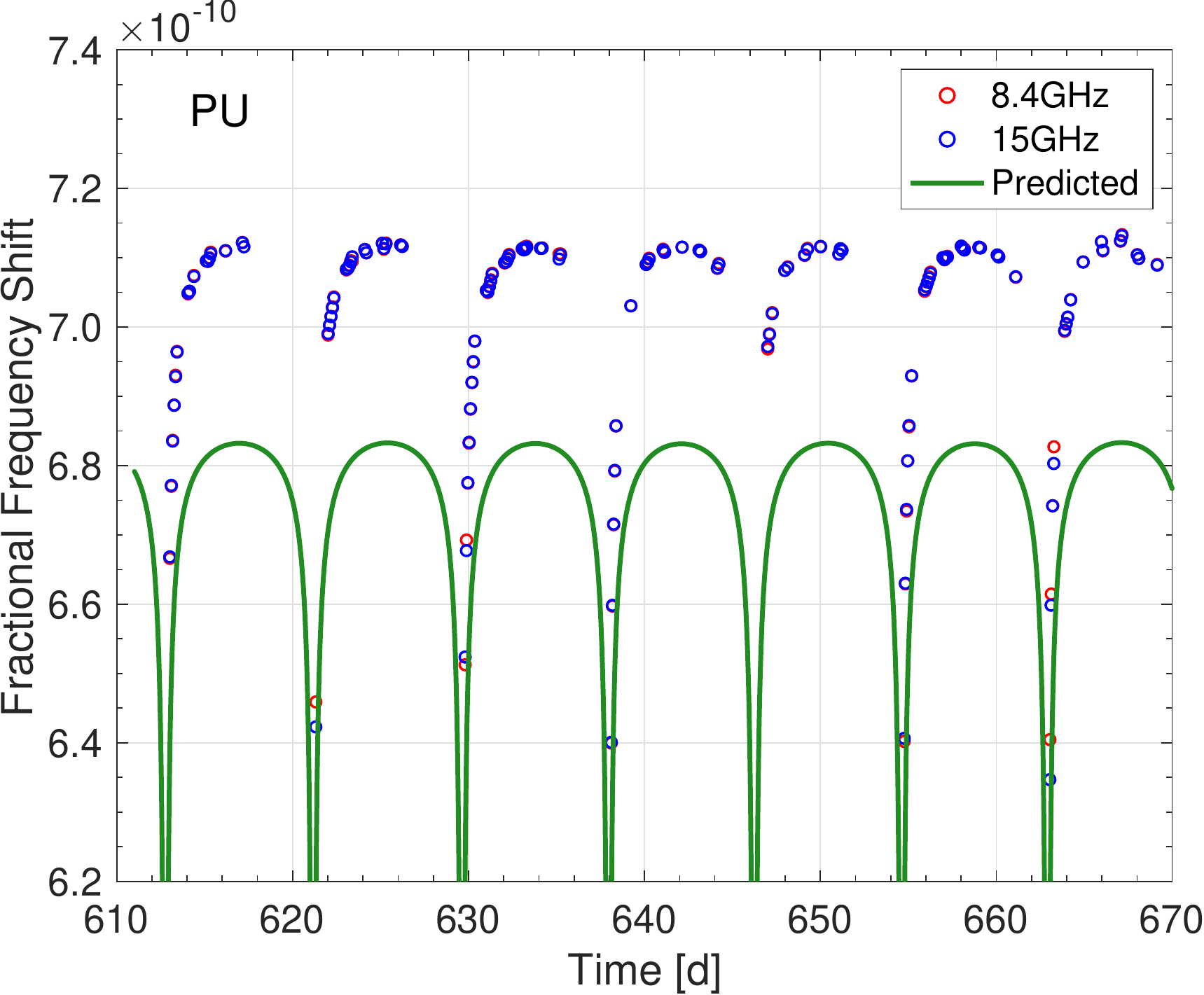}
	\caption{}
\end{subfigure}%
\begin{subfigure}{0.5\textwidth}
	\includegraphics*[width=6.5cm]{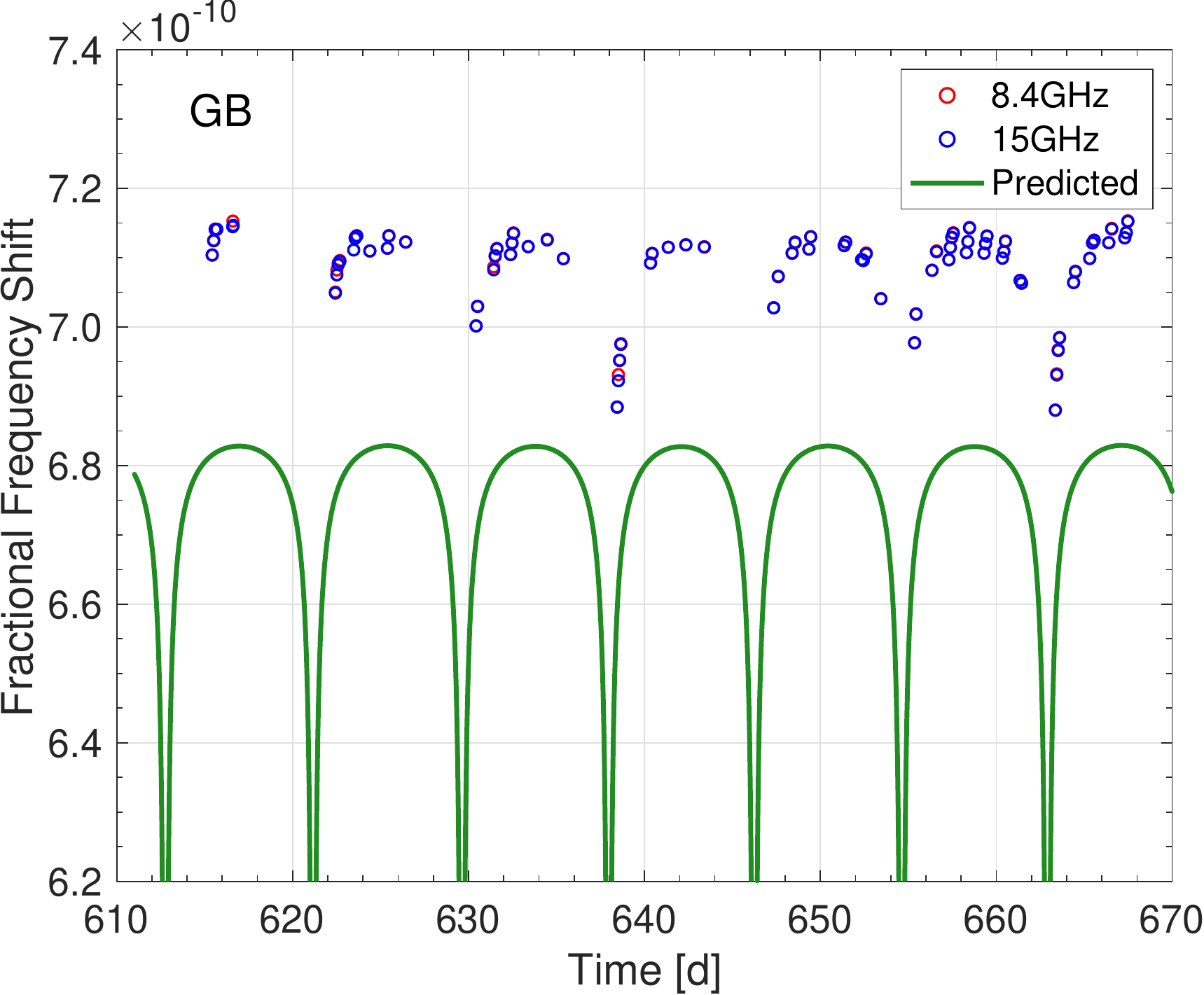}
	\caption{}
\end{subfigure}
\end{center}
\caption{Panels (a) and (b) show the observed biased gravitational redshift, $z_{\rm obs}^{\rm b}$, from each session at PU and GB between 2012 and 2017. For each session, one point is plotted for each of the two downlink carrier frequencies. Panels (c) and (d) zoom in on 60 d to show the variation in the gravitational redshift in detail. The predicted gravitational redshift, $z$, is shown in green for comparison. The offset between the two is mostly due to the clock offset between the onboard and station H-masers. Time is given in days since 2012 January 1. The periods where there are fewer data points correspond to the summer months when the number of experiments with RadioAstron dropped off due to constraints on spacecraft attitude limiting the visibility of radio sources of interest. During these months, most sessions were closer to perigee, resulting in smaller gravitational redshifts.  } 
\label{figure4}
\end{figure}

\begin{figure}[!t]
\begin{center}
\begin{subfigure}{0.5\textwidth}
	\includegraphics*[width=6.5cm]{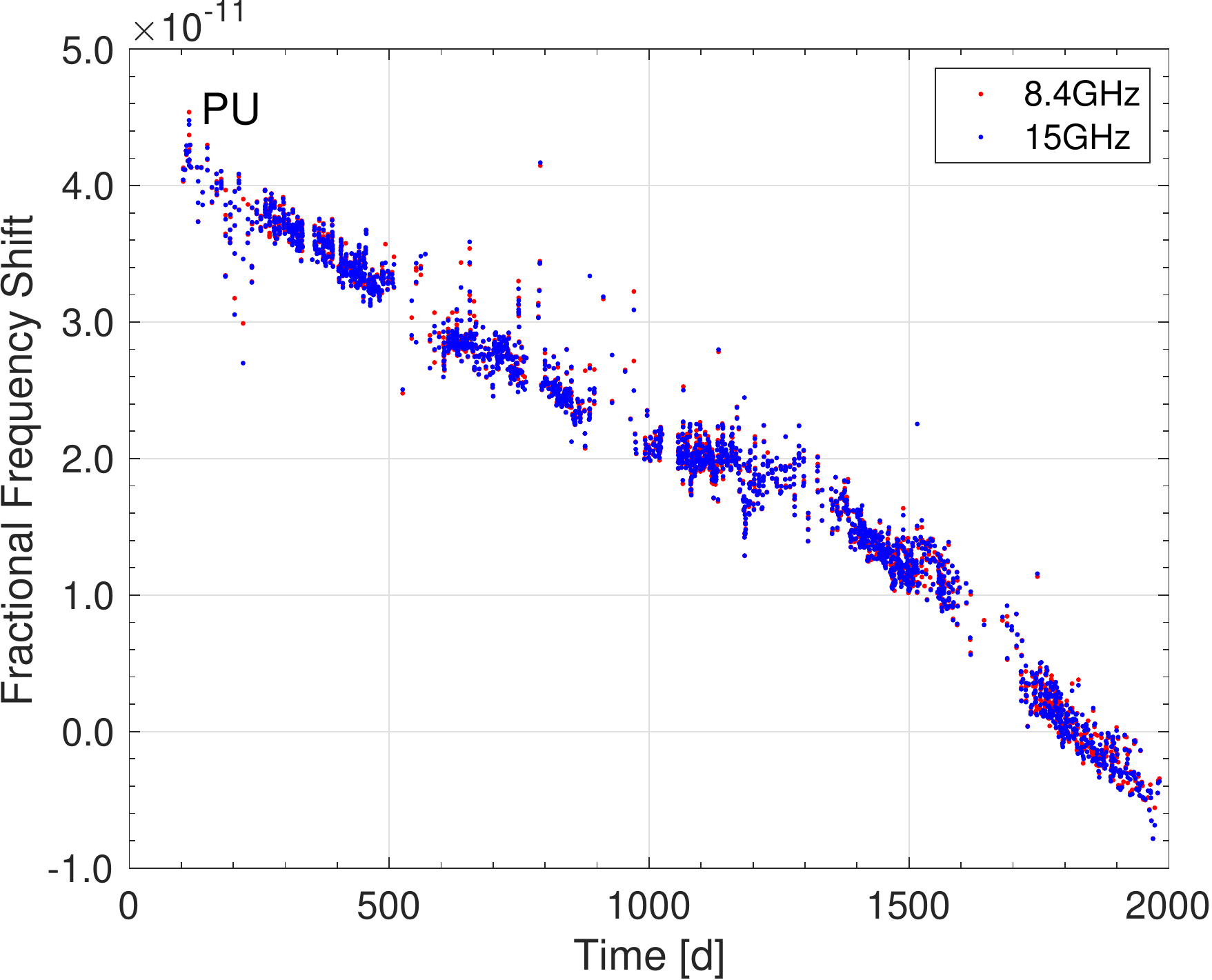}
	\caption{}
\end{subfigure}%
\begin{subfigure}{0.5\textwidth}
	\includegraphics*[width=6.5cm]{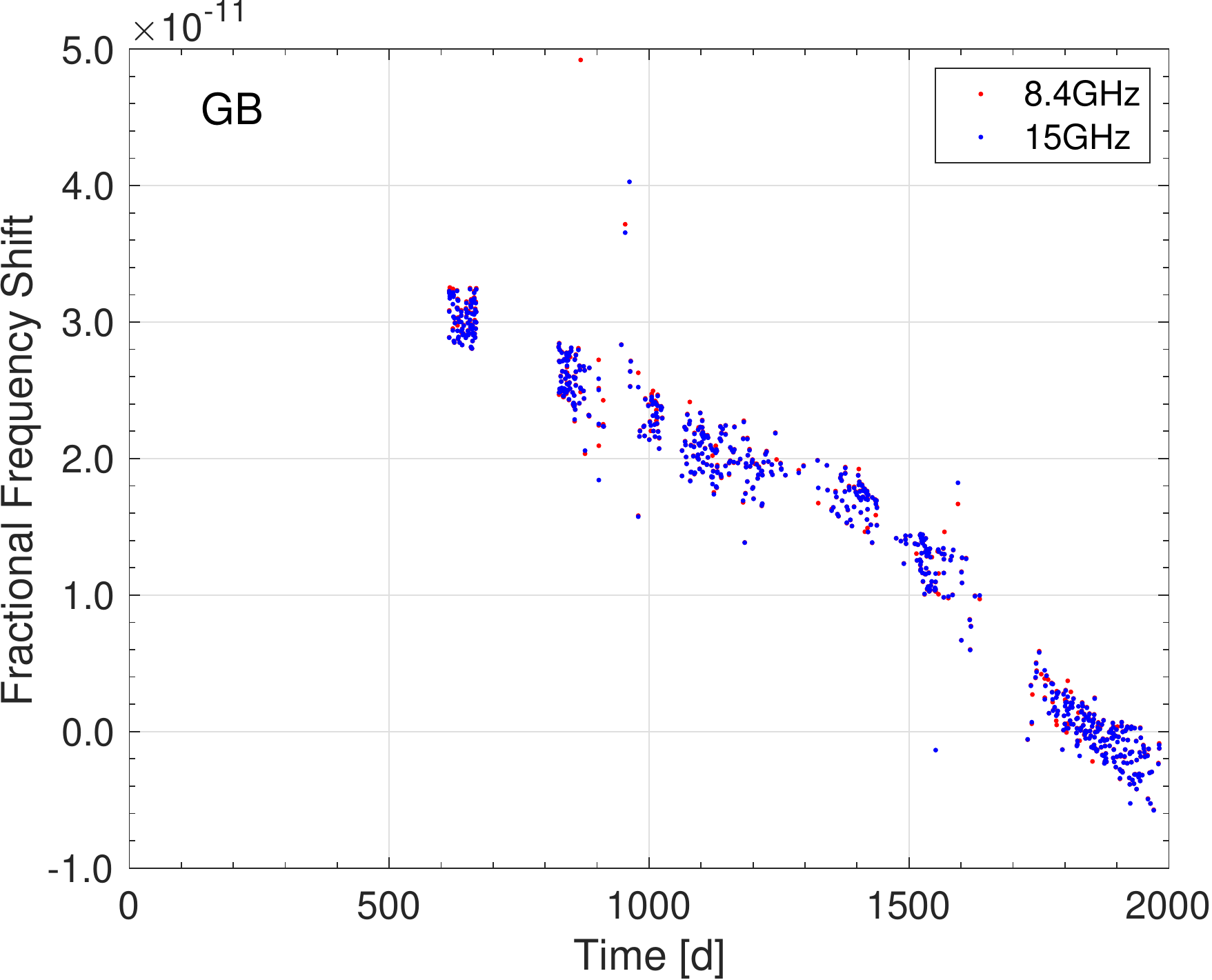}
	\caption{}
\end{subfigure}
\end{center}
\caption{Panel (a) and (b) show the observed biased gravitational redshift minus the predicted gravitational redshift from each session for PU and GB between 2012 and 2017. This difference is mostly due to the clock offset between the onboard and ground station H-masers drifting over time, apart from any contribution due to a possible non-zero violation parameter. Time is given in days since 2012 January 1. } 
\label{figure5}
\end{figure}

Our goal is to estimate $\epsilon$. Starting with the interpolated frequency measurement from each session, we obtain $z_{\rm obs}^{\rm b}$ by subtracting the Doppler effects computed using NOVAS \footnote{NOVAS is a software library for astrometry-related numerical computations provided by the United States Naval Observatory.  \url{http://aa.usno.navy.mil/software/novas/novas_info.php}} for the position and velocity vectors of the stations and the determined orbital state of the spacecraft. The geodetic coordinates of GB are known from VLBI observations to an accuracy more than sufficient for our purposes. However, the coordinates for PU are only known within a few meters which affects how accurately the non-relativistic Doppler shift can be computed and which, in turn, could have an impact on the estimate of $\epsilon$. In Figure \ref{figure4},  panels (a) and (b) show observed biased gravitational redshift measurements at PU and GB from 2012 to 2017, which are mostly due to the gravitational redshift including a possible violation due to $\epsilon \neq 0$ and the clock offset between the onboard and station H-masers. Panels (c) and (d) zoom in on 60 d over which the variation of the gravitational redshift is clearly visible and can be compared to the predicted value for $z$. The effect of the clock offset drifting over time is shown in Figure \ref{figure5}.

Starting with $N$ sessions, we formed pairs of sessions at times $t_i$ and $t_j$, where $i < j \leq N$ and $\Delta t = t_{\rm j} - t_{\rm i}$, over a maximum time interval, $\Delta t_{\rm max}$, where $0 < \Delta t \leq \Delta t_{\rm max}$. While a maximum of $N-1$ independent pairs are possible by pairing each of the N sessions with another session, the restriction on $\Delta t_{\rm max}$ results in fewer than $N-1$ pairs. The choice of which specific sessions to pair within the time interval is made to maximize the magnitude of the gravitational redshift difference, $|\Delta z|$, and thus maximize the sensitivity to a possible violation. By differencing the observed biased gravitational redshift to obtain $\Delta z_{\rm obs}^{\rm b}$, the constant clock offset, $y_0$, is cancelled. Fitting these differences to a model based on Equation (\ref{eq:5}) allows the violation parameter, $\epsilon$, to be determined along with $y_1$. However, if the mean of the ratio of $\Delta t$ and $\Delta z$ over time, $ \langle\Delta t / \Delta z\rangle$, is sufficiently small, which can be arranged by using a $\Delta t_{\rm max}$ that is relatively short, say $<10$ d, then the effect of the clock drift (as shown in Figure \ref{eq:5}) is small compared to the accuracy of $\epsilon$ achievable in our analysis and can be ignored.

\section{Estimate of $\epsilon$}

\begin{figure}[!t]
\begin{center}
\begin{subfigure}{0.5\textwidth}
	\includegraphics*[width=6.5cm]{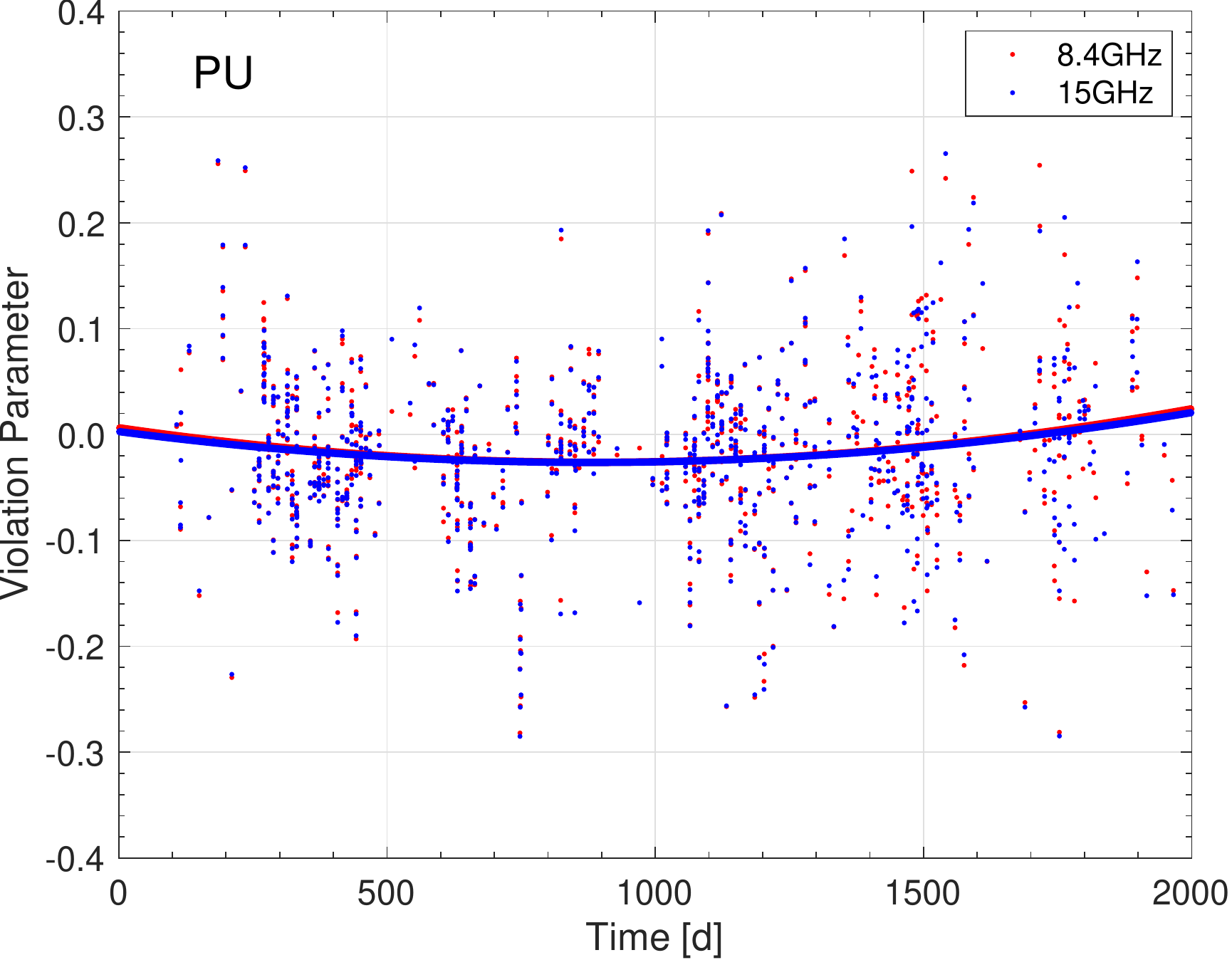}
	\caption{}
\end{subfigure}%
\begin{subfigure}{0.5\textwidth}
	\includegraphics*[width=6.5cm]{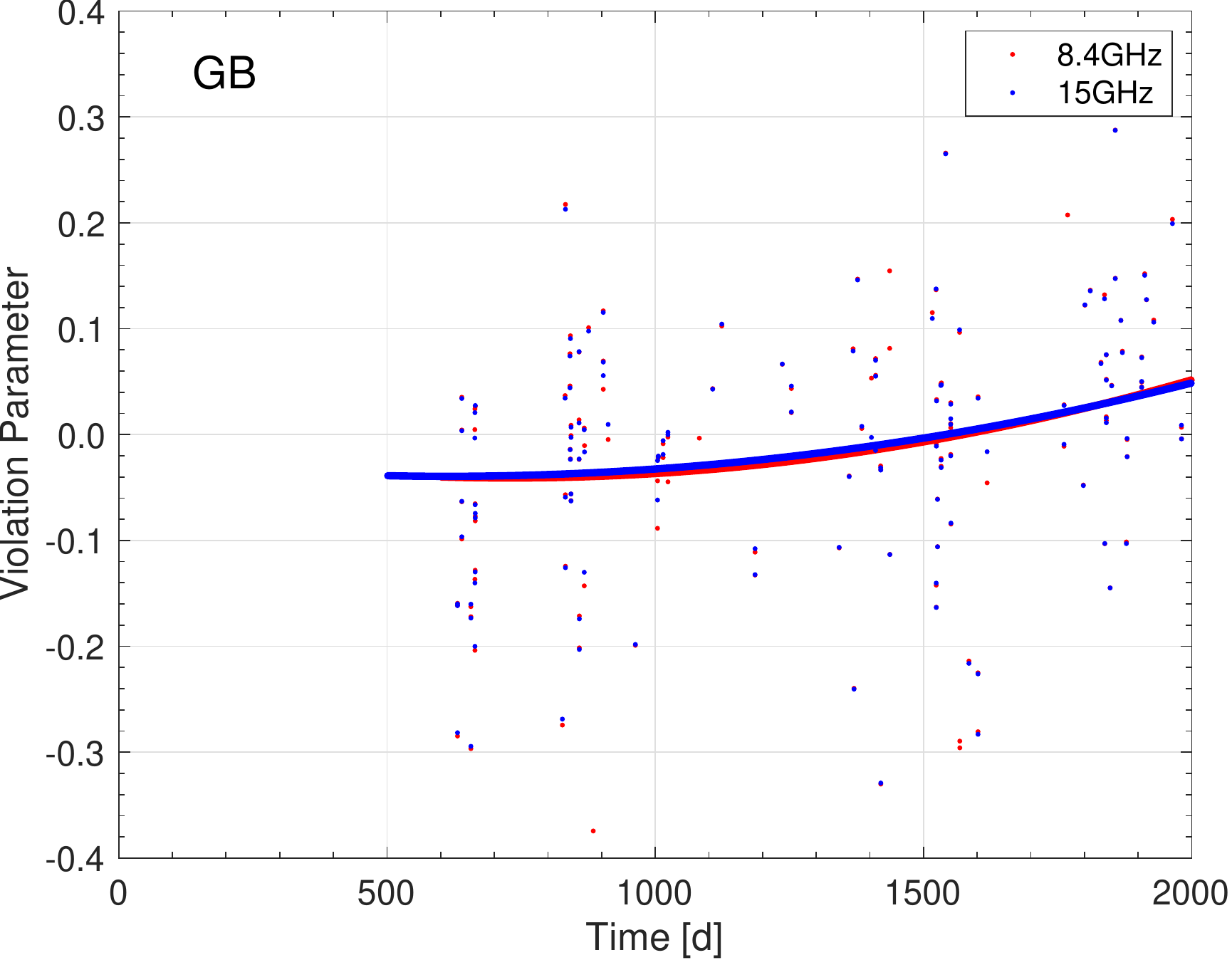}
	\caption{}
\end{subfigure}
\end{center}
\caption{Estimates of $\epsilon$ for each downlink carrier frequency computed using $\Delta z_{\rm obs}^{\rm b}$ from each pair of 1 h sessions where $| \Delta z|_{\rm min} = 1 \times 10^{-11}$ and $\Delta t_{\rm max} = 4.5$ d. Each point is plotted at the mid-point of the session pair. The rms scatter of 0.07 at PU and 0.08 at GB, is $\sim$ 3 times smaller than what would be expected from the uncertainty due to errors in the orbital velocity of the spacecraft found to be $1.36$ mm/s \citep{Zak18} in any direction. The parabolic fits, almost indistinguishable at the two downlink carrier frequencies, highlight a trend which we consider to be a systematic error. Time is given in days since 2012 January 1.} 
\label{figure6}
\end{figure}

For estimating $\epsilon$, we maximize the signal and minimize systematics and noise by considering only those session pairs with the largest gravitational redshift differences. We choose a minimum magnitude of the predicted gravitational redshift difference of $|\Delta z|_{\rm min} = 1 \times 10^{-11}$ over a time interval of less than $\Delta t_{\rm max} = 4.5$ d, about half an orbit, resulting in a weighted mean of $\langle\Delta t / \Delta z\rangle = 4 \times 10^{9}$ d for PU and GB. This choice allows the drift parameter, $y_1$, to be ignored but it still produces a sufficient number of pairs for our statistical analysis. The parameter $\epsilon$ is computed for each pair. Outliers are eliminated using a filter based on a $3\sigma$ criterion resulting in fewer than 6\% of pairs being excluded. In Table \ref{table1}, we list the mean values for our estimates of $\epsilon$, together with their standard errors, separately for PU and GB and for the two downlink signals. Individual estimates of $\epsilon$, for each session pair, are plotted in Figure \ref{figure6}. The solutions for the mean of $\epsilon$ at the two downlink carrier frequencies and for the two stations are all within a combined 0.5 $\sigma_{\rm stat}$. However, the solutions show a trend with time that we fit by a parabola to estimate systematic errors. The fit is very similar for the two downlink carrier frequencies at a given station and somewhat similar between the stations. This trend is expected to be due to unmodeled effects, possibly related to the error in accounting for the non-relativistic Doppler shift. We tried excluding sessions at times when the uncertainty in the orbital state parameters is highest, for example, during the summer months when there are fewer sessions as well as near perigee, but found the resulting fits to be within $1 \sigma$ of the combined statistical uncertainty. In fact, no subset of the data has been found that eliminates the systematic trend. We also analysed the effect of not accounting for the clock drift between the H-masers and, as expected, found that it does not explain the systematic trend. We estimate the magnitude of the systematic effect using the maxima and minima of the parabolas in the time range of the observations given in the table as $\sigma_{\rm syst,fit}$. Our final estimate of the systematic errors, $\sigma_{\rm syst}$, enlarges this range by taking into account the 68\% confidence bands of the fit.\footnote{Both linear and parabolic fits were tried giving systematic errors with 68\% confidence bands included of 0.017 to 0.028, respectively. We lean away from a cubic fit or higher as these aren't suggested by the scatter of the data and, instead, use a parabolic fit yielding a more conservative systematic error of 0.030 after rounding up.} Our average for $\epsilon$, weighted by the number of pairs, with statistical and systematic errors is:

\begin{table}[!t]
\begin{center}
\caption{Mean values for $\epsilon$ with uncertainties}
\label{table1}
\begin{tabular}{lrrrr}
\hline
                                         &\multicolumn{2}{c}{PU}  &\multicolumn{2}{c}{GB} \\
                                         &8.4 GHz      & 15 GHz      & 8.4 GHz      & 15 GHz \\
\hline
$\epsilon$                       & -0.018       & -0.018        &-0.006         & -0.002 \\
$\sigma_{\rm stat}$        & 0.002        & 0.002         & 0.005         & 0.005 \\
$\sigma_{\rm syst,fit}$   & 0.020        & 0.015         & 0.040          & 0.040 \\
$\sigma_{\rm syst}$       & 0.025        & 0.020         & 0.055          & 0.060 \\
No. pairs                          & 967           & 965             & 191              & 182    \\
\end{tabular}
\end{center}
\end{table}

$$\epsilon = -0.016 \pm 0.003_{\rm stat} \pm 0.030_{\rm syst}$$

\noindent This estimate of $\epsilon$ is consistent with zero within the combined uncertainties.

\section{Discussion}

We have presented an estimate of $\epsilon$ on the basis of 1-way frequency measurements at two stations and at two downlink carrier frequencies. Since the constant clock offsets between the onboard H-maser and the H-masers at the stations is difficult to measure independently of the gravitational redshift, only differenced frequency measurements are used to estimate $\epsilon$. Given our criteria for pairing observations, the difference in gravitational redshift is, on average, $\sim$ 30 times smaller than the absolute effect, and therefore the expected statistical noise floor is about 30 times larger. The statistical uncertainty for $\epsilon$ is approximately at the expected level and cannot be reduced much further using the 1-way data obtained at the ground stations. The systematic uncertainty, however, is tenfold larger. If the systematic trend in the fit of $\epsilon$ over the five years of observations were to be accounted for, the total error would be reduced significantly. This systematic effect is possibly due to the error in accounting for the non-relativistic Doppler shift, $-\dot{D} / c$, necessary when using 1-way data alone. Our dedicated interleaved observations of 1-way and 2-way data, however, offer the possibility of mostly eliminating the non-relativistic Doppler shift and potentially allow three orders of magnitude more accurate estimates \citep{Lit18}.

\section{Conclusions}
In summary:
\begin{itemize}
\item
For 5 yr a varying gravitational redshift was observed with the RadioAstron spacecraft, which is in an eccentric orbit from near Earth out to a distance of $\sim$ 350,000 km.
\item
The observations were made with the ground stations at Green Bank, USA, and Pushchino, Russia, in the 1-way operating mode at the downlink carrier frequencies of 8.4 and 15 GHz.
\item
These are the first measurements of the varying flow of time over such large distances in the vicinity of Earth.
\item
We estimate the violation parameter $\epsilon = -0.016 \pm 0.003_{ \rm stat} \pm 0.030_{ \rm syst}$
\item
This estimate of $\epsilon$ is consistent with zero within the combined statistical and systematic uncertainties and therefore consistent with the predictions of the Einstein Equivalence Principle which is foundational to general relativity.
\item
While the statistical uncertainty is at the expected level, the systematic uncertainty is tenfold larger, possibly due to the error in accounting for the non-relativistic Doppler shift.
\item
Dedicated interleaved observations using the 1-way and 2-way modes mostly eliminate the non-relativistic Doppler shift and promise to reduce the total uncertainty of $\epsilon$ by possibly three orders of magnitude. 
\end{itemize}

\section{Acknowledgements}
The RadioAstron project is led by the Astro Space Center of the Lebedev Physical Institute of the
Russian Academy of Sciences and the Lavochkin Scientific and Production Association under a contract with the Russian Federal Space Agency, in collaboration with partner organizations in Russia and other countries. The National Radio Astronomy Observatory is a facility of the National Science Foundation operated under cooperative agreement by Associated Universities, Inc.
The work of DAL, MVZ and VNR is supported by the RSF grant 17-12-01488 (estimation of error covariance matrix for RadioAstron gravitational redshift experiment). Research at York University was supported by a grant from the NSERC of Canada.

%%%%%%%%%%%%%%%%%%%%%%%%%%%%%%%%%%%%%%%%%%%%%%%%%%%%%%%%%%%%%%%%%%%%%%%%%%%%%
%% Appendices
% \appendix

\end{document}